\documentclass[prl,twocolumn,aps,10pt,superscriptaddress,notitlepage,nofootinbib]{revtex4}  
\usepackage{epsfig,amsmath,amssymb,amsfonts,graphicx,color,calc}
\usepackage{bbm}
\usepackage{graphicx}

\usepackage{cleveref}

\crefname{equation}{Eq.}{Eqs.}
\crefname{figure}{Fig.}{Figs.}

\let\a=\alpha  \let\g=\gamma

\let\s=\sigma \let\t=\tau

  \let\th=\theta

 \def\VV{{\cal V}}
\def\FF{{\cal F}}

\def\DD{{\cal D}}

\def\doubleunderline#1{\underline{\underline{#1}}}
\def\to{\rightarrow}

\newcommand{\beq}{\begin{equation}} \newcommand{\eeq}{\end{equation}}

\begin{document}

\title{Jamming in multilayer supervised learning models}
\author{Silvio Franz} 
\affiliation{LPTMS, Universit\'e Paris-Sud 11,
UMR 8626 CNRS, B\^at. 100, 91405 Orsay Cedex, France}

\author{Sungmin Hwang}
\email{sungmin.hwang@lptms.u-psud.fr}
\affiliation{LPTMS, Universit\'e Paris-Sud 11,
UMR 8626 CNRS, B\^at. 100, 91405 Orsay Cedex, France}

\author{Pierfrancesco Urbani}
\affiliation{Institut de physique th\'eorique, Universit\'e Paris Saclay, CNRS, CEA, F-91191 Gif-sur-Yvette}

\begin{abstract}

  Critical jamming transitions are characterized by an astonishing
  degree of universality. Analytic and numerical evidence points to
  the existence of a large universality class that encompasses finite
  and infinite dimensional spheres and continuous constraint
  satisfaction problems (CCSP) such as the non-convex perceptron and
  related models.  In this paper we investigate multilayer neural
  networks (MLNN) learning random associations as models for CCSP
  which could potentially define different jamming universality
  classes. As opposed to simple perceptrons and infinite dimensional
  spheres, which are described by a single effective field in
  terms of which the constraints appear to be one-dimensional, the
  description of MLNN, involves multiple fields, and the constraints
  acquire a multidimensional character. We first study the models
  numerically and show that similarly to the perceptron, whenever
  jamming is isostatic, the sphere universality class is recovered, we
  then write the exact mean-field equations for the models and
  identify a dimensional reduction mechanism that leads to a scaling
  regime identical to the one of infinite dimensional spheres. We
  suggest that this mechanism could be general enough to explain
  finite dimensional universality. 
\end{abstract}

\maketitle

Jamming is a common phenomenon that occurs in random systems subject
to constraints.  Examples can be found in different contexts such as
hard object packing \cite{ConwaySloane}, constraint satisfaction problems (CSP) in
computer science \cite{Tsang1993} and supervised learning \cite{Hastie2009}.  In these problems, one is
asked to find configurations that satisfy a
given set of constraints.  Increasing the hardness of the problem by either
increasing the number of constraints or making them more stringent,
one can find the point -possibly dependent on the search protocol-
above which solutions no longer can be found.  When
this point is identified, we say the system undergoes the jamming
transition \cite{LN10,Behringer2018,PZ10}.

In the packing problem of hard spheres, the task is to find arrangements of $N$ spheres in a
$d$-dimensional volume $V$ in such a way they do not overlap
\cite{ConwaySloane, PZ10}.  Starting from a low density configuration
of non-overlapping spheres, the system is compressed until it jams
forming a rigid contact network \cite{LS90}.  Jammed packing reached
by different compression protocols are critical and share
universal features.  They are generically {\it isostatic}
configurations of spheres that display marginal mechanical stability
\cite{TW99, LN98, OSLN03, WSNW05, LN10}.  
Their statistical properties
can be described by a set of power laws characterized by critical
exponents that satisfy simple scaling relations \cite{MW15,
  FPSUZ17,Wy12}, and in the limit of infinite dimension, they
can be computed exactly \cite{CKPUZ13, CKPUZ17}.  It turns out that
these mean-field predictions work quantitatively well even in
three and higher dimensional packing \cite{LDW13,
  CCPZ15}.

In computer science, discrete CSP are important  as they represent the building
blocks of the theory of computational complexity \cite{AB09}.
Highlighted by the recent success on sphere packing \cite{FP15,
  FPUZ15, FPSUZ17}, it has been shown that CSPs involving continuous
variables (CCSP), while undergoing a phenomenology similar to the
discrete counterpart, can show the additional interesting feature of
critical jamming\cite{FP15,
  FPSUZ17}. 

The third example of CCSP is provided by supervised learning of
classification tasks in neural networks. In this case one needs to find
the network parameters such that all the examples contained in the training
dataset are correctly classified \cite{EVbook}. 

If the classification task is simple for a given architecture, there is no limit to the number of
examples that can be learned. On the other hand, if the classification
rule is too difficult for the network, a SAT/UNSAT transition appears as
the limit of the number of examples that can be correctly
classified. The extreme case of this setting is the one in which the dataset is
given by random patterns. 

The simplest model in this setting is the perceptron neural
net.  It consists in a single layer of input units directly connected
to a single output node \cite{EVbook, DG88}.  When used with random
data, it displays a jamming rigidity transition from a liquid phase
where jammed configurations are hypostatic and non-critical.  However,
the model has been recently twisted into a non-convex CCSP in
\cite{FP15}.  In this case, the SAT/UNSAT transition appears to be
isostatic, thermodynamically marginal and in the same universality
class of spheres \cite{FP15, FPSUZ17}. The same is observed in some
generalizations of the non-convex perceptron that include multibody
interactions \cite{Yoshino2018}. 

The nature of jamming points appears to be related to the presence,
or the absence of two kinds of marginality: isostaticity, the marginal mechanical stability
that can develop at jamming, and the marginal thermodynamical
stability of the phase where jamming occurs. Both kind of marginality
are present in high dimensional spheres and in all other models where
critical power laws have been found. If any of these two marginalities is absent,
one does not find critical power laws \cite{FP15}.  
Therefore, given both kind of marginality,
it is natural to ask how many universality classes can arise and how to compute the 
associated critical exponents.
Up to now only one universality class has been found which coincides with
jamming of hard spheres in infinite dimension.

Since it does not seem to us conceivably possible to tackle with the
problem in full generality, we consider in this paper the case of
multilayer neural networks. We concentrate on the simplest multilayer
generalizations of the perceptron, which still allow for analytic
treatment and could in principle define new universality classes, the
parity and the committee machines and some variations of them.  The
reason to believe that these models could define new universality
classes is technical. Thermodynamics marginality of the aforementioned
kind occurs in marginal glass phases, described by continuous -or
full- Parisi replica symmetry breaking (fRSB).  We show below that for
a number of hidden units $K$, the flow equations describing isostatic
jamming points are $K$-dimensional generalizations of the 1D partial
differential equations that hold for the spheres and the perceptron.
The scaling solutions of this type of partial differential equations
are known to be strongly dependent on dimensionality and this could
lead to different universality classes.  The main result of this paper
is that, despite their multi-dimensional description, critical jamming
of multilayer networks still lies in the sphere packing universality
class. In what follows we first present compelling numerical evidence
that in a variety of models the critical exponents coincide with the
ones of the spheres, then we sketch a deep dimensional-reduction
mechanism valid in the scaling regime close to jamming. The argument
is fully detailed in the SM.

\paragraph{Definition}
Let us focus on the simplest feed-forward neural networks for binary
classification with two layers.
A network has $N$ input units, a single output unit and a hidden layer with $K$
units. We consider the case where the weights of the input-to-hidden layer are
learnable while the mapping from hidden to output is fixed, and
consider the case of large -and eventually infinite- $N$ for finite -and
small- $K$. Specifically
given an activation pattern $\underline \xi\equiv\{\xi_1,\ldots , \xi_N\}$ in input, for
given weights $\doubleunderline w\equiv \{w_{i,j}\}_{i=1,\ldots, K; j=1,\ldots, N}$,
one defines $K$ effective fields $\underline h\equiv\{h_1,\ldots , h_K\}$
\begin{eqnarray}
  \label{eq:1}
  h_i\left[\doubleunderline w,\underline \xi\right]= \sum_{j\in {\cal L}_i} w_{i,j}\xi_j,
\end{eqnarray}
where ${\cal L}_i\subset \{1,...,N\}$ is the `receptive field' of unit
$i$.  The output of the network is then computed according to some binary
function of the fields ${\rm sgn } (\FF({\underline h}))$.
Note that due to the arbitrariness of the choice of $ \FF({\underline
  h}) $, our formalism covers a 
very broad class of CCSP.
For simplicity, we 
consider the case of disjoint receptive fields
with $n=N/K$ elements. 
Additionally, while our scaling theory holds in general, we concentrate on three notable machines
{
	\medmuskip=-1mu
\thinmuskip=-1mu
\thickmuskip=-1mu
\scriptspace=0pt
\begin{align}
\FF[{\underline h}]=\left\{\begin{array}{ll}
\prod_{i=1}^K h_i &\textrm{{\bf parity}}\\      
\sum_{i=1}^K {\rm erf\;} h_i &\textrm{{\bf soft committee}}\\
\frac 1K\sum_{i=1}^K \rho_{\rm ReLU}(h_i, \sigma) -\sigma &\textrm{{\bf ReLU 2-layer}}.\\
\end{array}\right.
\end{align}
}
The \emph{ReLU 2-layer} machine is composed of Rectified Linear hidden Units,
\beq
\rho_{\rm ReLU}(h_i, \s) = \left( h_i - \sigma \right) \th\left(h_i - \sigma \right)
\eeq
being $\th(x)$ the Heaviside step function and $\sigma>0$ an activation threshold.

We consider the problem of random associations, where $M$ random binary or
Gaussian $N$-component vectors ${\underline \xi}^\mu\equiv\{\xi^\mu_1, \ldots, \xi^\mu_N\}$ are associated to
binary labels $\tau^\mu$ taking values $\pm 1$ with equal
probability. 
Then, one seeks for an assignment of $\doubleunderline w$
such that for all $\mu=1,...,M=\a N$
\begin{eqnarray}
\label{eq:2}
\tau^\mu ={\rm sgn} \left( \FF\left[{\underline h}\left[\doubleunderline w,\underline \xi^\mu\right]\right] \right)
\end{eqnarray}
In order to add stability against noise, 
it is customary to demand that the fields ${\underline h}^\mu={\underline h}\left[\doubleunderline w,\underline\xi^\mu\right]$ lie far enough from the
border $\tau^\mu \FF\left[{\underline
    h}^\mu \right] =0$.
One can then introduce 'gap variables', $ \Delta_\mu\equiv
\Delta[{\underline h}^\mu]$ which are required to be positive in SAT
assignments of ${\doubleunderline w}$. In the parity and committee machines one defines
\begin{eqnarray}
  \Delta_\mu\equiv \Delta[{\underline h}^\mu]= \tau^\mu \FF[{\underline h}^\mu]-\sigma,
\label{gap_parity}
\end{eqnarray}
and in the ReLU
\beq
\Delta_\mu=\tau_\mu\left[\frac 1K \sum_{i=1}^K \rho_{\rm ReLU}(h_i, \sigma) -\sigma'\right]\:.
\label{gap_ReLU}
\eeq where for simplicity we choose $\sigma'=\sigma$.  In order to
prevent overflow, we normalize the weights according to
$\sum_{j \in {\cal L}_i } w_{ij}^2=1.$ for all $i$. In the
case of the parity and committee machine, the
$\tau_\mu$ can be reabsorbed into the vectors
$\xi_\mu$ without affecting their statistical properties and we can
directly pose $\tau_\mu=1$. 
Even if this property does not hold for
the ReLU Machine we still set $\t_\mu=1$ in this last case
we train the network providing  examples from  one class only.

We also found instructive to study a variant of the model considered
in \cite{FP15}, a simple perceptron, where the 
 the input patterns are denoted by $\underline \xi^{\nu,A}$ 
are correlated according to
$\overline{\xi^{A,\nu}_i \xi^{A',\nu'}_j} = \delta_{ij}\delta_{\nu \nu'} \DD_{AA'} $
being $\DD_{AA'}$, a given correlation matrix.
For the simplest nontrivial case $K=2$, it is parametrized by a correlation parameter $\rho$ with 
$ \DD_{AA'} = (1-\rho)\delta_{AA'}+\rho $.
Despite the inherently different origin, the resulting replica approach leads to a similar $K$-dimensional Parisi flow equations \cite{MPV87}.

\paragraph{Numerical Simulations}
Learning of the weights matrix $\doubleunderline w$ of all these
models is performed by minimizing an empirical loss function. 
We choose the quadratic loss
\beq
H\left[\doubleunderline w; \{\underline \xi^\mu\} \right] =  \frac 12 \sum_{\mu=1}^M {\Delta_\mu}^2 \theta(-\Delta_\mu)\:.
\label{cost_function}
\eeq
which is zero iff all the gaps are positive. 

We are interested in understanding the properties of jammed
configurations of $\doubleunderline w$. We produced jammed
configurations by a 'decompression algorithm' similar to the one used
in studies of jamming of spheres \cite{OSLN03}.  For a given value of
$\alpha$, we initialize the network in a highly
\emph{under}-parametrized UNSAT regime at large $\sigma$. We minimize
the loss function (\ref{cost_function}) using the routine provided in
DLIB library, implemented based on L-BFGS algorithm \cite{DLIB09}.
Then we reduce $\s$ till we reach a jamming point where the value of
the loss is zero. We find rather small finite size effects in the
parity and committee machines allowing the precise determination of
the the transition point. The case of the ReLU machine turns out to be
slightly more difficult due to systematic finite-size effects  (See SM).
Coherently with the theoretical analysis (see SM sec.4, fig. 5), we
find that  jamming  is always \emph{isostatic} for all the values of  $\alpha$
that we tested in the Parity and ReLU machines. In the committee case,
we find instead isostaticity only for $\alpha$ beyond $\alpha_{RS}\approx 1.75$ and
hypostaticity below this value.  This is similar to what happens in the perceptron
model, where following the jamming line there is a phase transition point
from a replica symmetric hypostatic regime at lo $\alpha$ to a fRSB
isostatic one at large $\alpha$.  As stated in the introduction we concentrate here on the
critical RSB case. 

Close to the jamming transition we measure the following 
observables:
i) \underline{contacts}, the number $z$ of unsatisfied constraints
normalized by $N$, 
ii) the pdf of positive gaps at jamming. 
iii) \underline{force distribution}, measured as the pdf of 
negative gaps scaled by their mean, in the UNSAT phase close to
jamming \cite{OSLN03}. 
\begin{figure}[t]
\begin{center}
\includegraphics[width=0.9 \columnwidth]{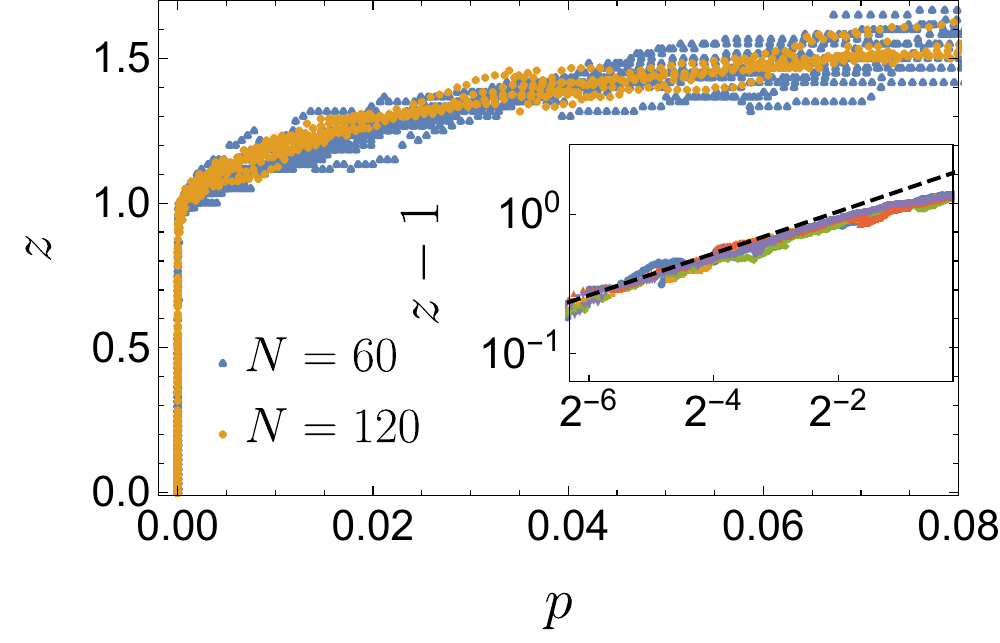}
\caption{Contact number as a function of the pressure in the Parity machine for $K=2$ and $\alpha =3$ with two different system sizes (10 realizations for each size).
Increasing the system size, the fluctuations decrease. 
Close to unjamming, we find $z-1\sim p^{1/2}$ as is shown in the inset where the same plot is shown in log scale.
}
\label{fig:Contacts}
\end{center}
\end{figure}

In Fig.~\ref{fig:Contacts} we plot the contact number as a function of the pressure for the Parity machine with $K=2$ and two different system sizes.
The pressure is defined as the average of negative gaps.
We find that at jamming, when the pressure vanishes, $z$ approaches
the \emph{isostatic} value $z=1$.  Furthermore, the inset shows that
as in spheres, $z-1\sim p^{1/2}$ close to unjamming \cite{OSLN03,
  WSNW05, FPUZ15}.

\begin{figure}[t]
\centering
\includegraphics[width=0.7\columnwidth]{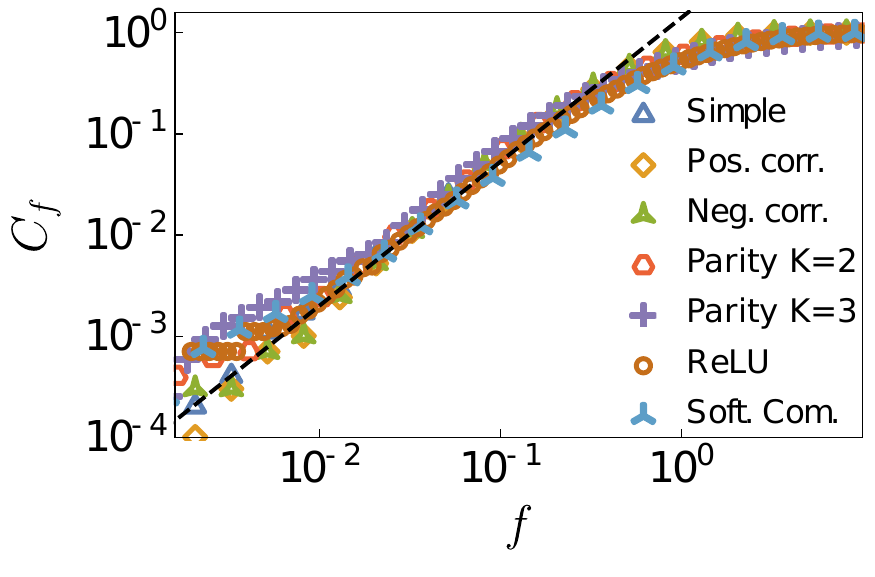}
\includegraphics[width=0.7\columnwidth]{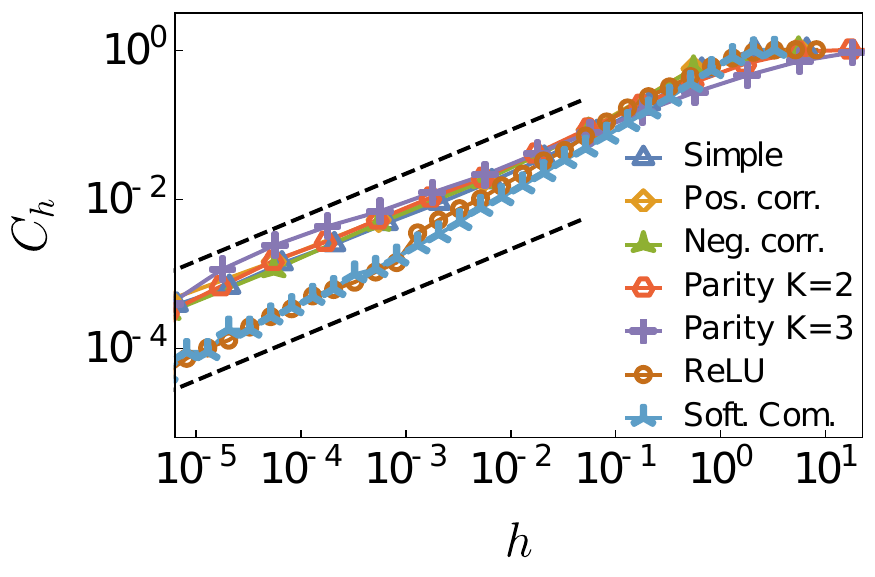}
\caption{Cumulative distributions of forces and gaps near a jamming point at $\alpha =3$. 
The simple perceptron corresponds to the model studied in \cite{FP15, FPUZ15, FPSUZ17}.
In agreement with the replica analysis, both distributions follow power-law with exponents $C_f \sim f^{1+\theta}$ and $C_h \sim h^{1-\gamma}$, respectively. 
The force and gap variables are rescaled such that their power-law regions are collapsed into a dashed line, indicating the theoretical slope of power-law exponents.
}
\label{fig:GapDistribution}
\end{figure}
In Fig. \ref{fig:GapDistribution}, we have shown the CDF of the force
$C_f$ and gap $C_h$ distributions as defined above for all models
under study. 
We observe that the distributions follow a power-law behavior in the limit $f \to 0^+$  and $h \to 0^+$ that is $C_f \sim f^{\th +1}$ and $C_h\sim h^{1-\g}$ with $\th\simeq 0.42$ and $\g\simeq 0.41$.
Surprisingly, these power-law exponents are shown to be  universal for all tested models and coincide (within our numerical precision) with the ones obtained for hard spheres in infinite dimension \cite{CKPUZ14}.

Furthermore we note that for the soft committee and ReLU machines, the universal power-law regime seems to be preceded by a pre-asymptotic behavior. 
This is visible in the gap distribution where in the range of $\overline h \in [10^{-3}, 10^{0}]$ we have $C_h\sim h^{1+\th'}$ with $\th'\sim 0.2$. 

\paragraph*{Analytical approach}
The models that we have considered can be solved analytically through the replica method.
Here we will not present a complete derivation of the saddle point equations since 
the general framework is well established as can be found for example in \cite{FPSUZ17}.
The starting point of the analysis is to compute the logarithm of the Gardner volume given by
\beq
s = \frac{1}{N} \overline{\ln{\cal V} } 
=\frac{1}{N} \overline{\ln \int \DD\doubleunderline{w} 
	 \prod_{\mu=1}^M \theta(\Delta_\mu)}.
\label{free_entropy}
\eeq
The computation of the average over the training set (the overline in
Eq.~(\ref{free_entropy}))  
can be done with the replica trick
$
\overline{\ln \VV} =\lim_{m \to 0 }\partial_m \overline{\VV^m}\:
$
using the Parisi RSB ansatz to evaluate the saddle point over replica
matrices \cite{MPV87}. 

The phase diagrams of the Parity and committee machines have been
studied in the RS and 1RSB approximation \cite{BHK90,BHS92}.
Depending on the model and the parameters, jamming can occur in a RS
or RSB region of the phase diagram. In the parity machine, RSB is
needed in the entire range of $\sigma$.  The actual calculations have
been done in the 1RSB setting, however we believe that 1RSB is
unstable close to jamming, which occurs in a ``Gardner'' marginal
glass phase, and has an isostatic character. In the soft committee and
ReLU we have analysed the character of the jamming transition by
comparing the RS line of limit of capacity with the deAlmeida-Thouless
line of the replica symmetric solution becomes unstable (see SM).  It
turns out that in the ReLU, jamming always lies in a continuous RSB
region, while in the soft-committee, part of the jamming line is RS
and part fRSB.

Let us now sketch the analysis of the fRSB solution close
to jamming and discuss the mechanism by which universal scaling
emerges. As usual, the Parisi order parameter, the distribution of the
possible overlaps between SAT configurations of weights, 
is parameterized in terms of a continuous function $x(q)$ in some interval
$[q_m,q_M]$. The jamming point, where the Gardner volumes shrinks to a
point, is identified by the condition that the maximum overlap $q_M$
tends to one.   The peculiarity of the models we consider, is that 
 while in the high dimensional spheres \cite{CKPUZ13},
the perceptron \cite{FPSUZ17} and its generalizations \cite{Yoshino2018} the
gap variables depend on a single effective field -and gaps and fields
can be identified-,  here each gap
variable depends on $K$ different fields. As a consequence
the determination of the free-energy
requires the joint distribution of $K$ effective fields (one
for each hidden unit) $P(q_M, h_1,...,h_K)$, which is obtained as the terminal
point at $q=q_{M}$ of the solution of  multidimensional Parisi flow equations
for the 
two functions $P(q,\underline h)$ and  $f(q, \underline h)$:
\beq
\begin{split}
\frac{\partial f}{\partial q} &= -\frac 12  \left(\nabla^2_h f + x(q)  \left| \nabla_h f \right|^2 \right)\\
\frac{\partial P}{\partial q} &= \frac 12  \left(\nabla_h^2 P- 2x(q)  \nabla_h\cdot\left(P\nabla_h f\right) \right)
\end{split}
\label{multi_full}
\eeq
whose boundary conditions are
\beq
\begin{split}
f(q_M,\underline h) &= 
\int d \underline{h} \frac{e^{-\frac{|\underline{h}|^2}{2(1-q_M)}}}{(2\pi(1-q_M))^{K/2}}
 \theta(\Delta(\underline{h}))\\
P(q_m,\underline h) &= \frac{e^{-\frac{|\underline{h}|^2}{2 q_m}}}{(2\pi q_m)^{K/2}}.
\end{split}
\eeq These equations are complemented by a self-consistent equation
for $x(q)$, detailed in the SM. In the one dimensional case, close to
jamming, one finds a scaling regime where gaps tend to zero as the
jamming limit is approached, $h\sim\sqrt{1-q}$ for $q\to 1$, such that
$P(h,q) \approx {\mathcal{P}(\frac{h}{\sqrt{1-q}})}$. In our
multidimensional setting, fields do not have reasons to be small and
different scaling variables need to be identified. Close to jamming,
the relevant configurations of fields ${\underline h}$ should lie
close to the border of the allowed region
${\cal R}=\{ {\underline h}|\Delta {(\underline h)}>0\}$. Close to a
point $B$ on the border of ${\cal R}$, which we can parameterize
$ {\underline h}$ by the coordinate along the tangent plane on $B$,
$\delta {\underline h}_\parallel$ and the distance from it $\delta h_\perp$. Consistently we
can assume a scaling regime $\delta h_\perp\sim \sqrt{1-q}$ with a singular
dependence on $\delta h_\perp$, while the dependence on
$\delta{\underline h_{\parallel}}$ would remain regular. In this regime,
$\delta h_\perp$ and $\delta{\underline h_{\parallel}}$ become independent, and the
distribution of $P(q, {\underline h})$ factorizes in a singular part
independent of the choice of $B$, 
$\frac{1}{\sqrt{1-q}}P_{sing}\left(\frac{\delta
    h_\perp}{\sqrt{1-q}}\right)$, 
and a regular part $P_{reg}(1, \delta {\underline h_{\parallel}}|B)$.  This ansatz actually
solves the equations if a scaling regime is assumed, moreover, one
finds the same set of equations of the case of high dimensional
spheres. The behavior of force and gap distributions near jamming is
fully determined by the singular part, yielding the same power-law
exponents $\theta$ and $\gamma$ (see SM for details).

Finally, let us comment that while strictly speaking the limit of
capacity and its associated jamming transition only exist
thermodynamically for `unlearnable' rules, jamming phenomenon 
can also be relevant for metastable states in the case of 
learnable rules such as in a Teacher-Student
scenario \cite{EVbook} where examples are provided 
by a `teacher machine'. If the system undergoes a first-order phase transition from
a learning-by-heart phase to a generalization phase, the system makes
a transition only after the learning-by-heart phase becomes unstable
(a kind of Kirkwood instability \cite{FP87, FP99, AFUZ18}).  These
glassy states may jam eventually and therefore one expects the same
critical exponents of the purely random model (See SM for the complete
discussion).

\paragraph*{Conclusions}
In this work we have considered multilayer supervised learning models as models for constraint satisfaction problems with continuous variables.
The SAT/UNSAT transition corresponds to the limit of capacity for learning random input-output associations.
We have shown through numerical simulations that close to this point these models display the same critical behavior of hard spheres close to their jamming transition.
Based on these observations, we have developed a replica approach showing that when a fullRSB marginal phase is realized close to the SAT/UNSAT threshold.
Furthermore we have developed a scaling theory
showing how the hard spheres universality class can be recovered in these models.
By establishing the dimensional reduction mechanism, we have shown that the hard-sphere universality covers
a large class of supervised learning models. 
It is a great future project to finish the complete characterization of this universality across different types of CCSP.

\subsection*{Acknowledgments}
We thank G. Biroli, G. Parisi, J. Rocchi and M. Wyart for very useful
discussions and exchanges.  This work is supported by
``Investissements d'Avenir'' LabEx PALM (ANR-10-LABX-0039-PALM).
S. Franz and S. Hwang are supported by a grant from the Simons
Foundation (No. 454941, Silvio Franz).  

In an interesting paper
\cite{Geiger2018} that appeared at the same time of ours, multilayer
networks with fully learnable weights have been analyzed in a jamming
perspective. In that case hypostaticity is found and the presented
data suggests a different universality class. It would be interesting
to compare these finding with the single layer convex perceptron and
the solvable soft committee in the replica symmetric, hypostatic
regime.

\bibliography{HS}

\end{document}


\title{Supplementary Material for\\ ``Jamming in multilayer supervised learning models''}

\author{{ 
		Silvio Franz$^1$, Sungmin Hwang$^1$ and Pierfrancesco Urbani$^2$}\\
{\small \em $^1 $LPTMS, Universit\'e Paris-Sud 11,
	UMR 8626 CNRS, B\^at. 100, 91405 Orsay Cedex, France\\
$^2$Institut de physique th\'eorique, Universit\'e Paris Saclay, CNRS, CEA, F-91191 Gif-sur-Yvette\\
}}

\date{\today}
\maketitle

\section{Basic formalism}
The replica approach to the multilayer models can be
developed along the lines of \cite{FPSUZ17}.
In terms of $K$ effective fields ${\underline h}\left[\doubleunderline w,\underline\xi^\mu\right] )$ (Eq. (1) of the main text) for a given weight $\doubleunderline w$ and for each pattern $\underline\xi^\mu$, 
we begin by establishing the formal expression for the replicated expression of the Gardner volume:
\begin{align}
\overline{\VV^m} 
= \overline{\int \prod_{a=1}^{m} \DD \doubleunderline{w}^a 
	\prod_{\mu=1}^M \theta(\Delta( {\underline h}\left[\doubleunderline w^a,\underline\xi^\mu\right] ))},
\end{align}
where the overline denotes the average over the random patterns and $\DD\doubleunderline{w}$ is the flat measure over the $N$-dimensional sphere $\sum_{j \in {\cal L}_i } w_{ij}^2=1$ for each $i$.
This equation simply computes the volume of $m$-tuples of solutions such that $ \Delta( {\underline h}\left[\doubleunderline w^a,\underline\xi^\mu\right] )) > 0 $
for all $\mu = 1, \cdots, M$ of the given model encoded by the gap $\Delta(\underline{h})$.

This quantity may be evaluated based on a series of observations:
\begin{enumerate}
	\item The gap $\Delta(\underline{h}^a)$ depends only explicitly on the $K$ effective fields $\underline{h}^a$, thus it is crucial to understand the statistical properties of them.
	\item The $K$-fields $\underline{h}^a$, being the sum of large number of random variables, are Gaussian random variables for a fixed weight $ \doubleunderline{w}^a  $.
	\item The correlation $Q_{ab}^i$ of $i$-th field between a pair of replica $(a,b)$ is solely determined by the relative position of their weights, i.e., $Q^i_{ab} = \sum_{j \in {\cal L}_i } w^a_{ij} w^b_{ij}  $.
	\item The transformation from  $\doubleunderline{w}^a $ to $Q^i_{ab}$ yields the Jacobian $\exp(\frac{N}2 \log\det  Q^i)$ with $Q^i_{aa}=1$ due to the spherical constraint.
\end{enumerate}
Incorporating these observations, the replicated volume reads
\begin{align}
\overline{\VV^m} \propto  \int \prod_{i=1}^{K} \prod_{a<b}\de Q^i_{ab} e^{N\SS(Q^i)},
\end{align}
where
\begin{align}
\SS(Q^i) = \frac 12 \sum_{i=1}^{K}  \ln \det Q^i + \a\ln \ZZ.
\end{align}
The effective single variable partition function $\ZZ$ is then given as a Gaussian average:
\begin{align}
\ZZ = \Avr{\prod_{c=1}^m \theta( \Delta(\underline{h}^c) )},
\end{align}
where $ \Avr{\cdot} $ stands for the corresponding Gaussian average with the correlation given by $\Avr{h_i^a h_j^b} = \delta_{ij} Q_{ab}^i$.

The saddle point equations for the overlap matrices $Q_{ab}^{i}$ can be obtained by setting up a fRSB hierarchical ansatz \cite{MPV87}. 
Within this setting, the off-diagonal elements of each $Q_{ab}^i$ are parametrized by $q_i(x)$ in the range $x\in[0,1]$.
Additionally, the statistical symmetry under the exchange of hidden units implies that the solution is of the form $ q_i(x)=q(x) $.
These equations provide a simpler form if it is written in terms of its inverse $x(q)$, 
\begin{align}
\frac{q_m}{\l^2(q_m)}+\int_{q_m}^q \de p \frac{1}{\l^2(p)} &=\a \int_{-\infty}^\infty\de \underline h P(q,\underline h )  |\nabla_h f(q,\underline h)|^2  ,
\label{full_correlated}
\end{align}
where $q_m = q(0)$, $q_M = q(1)$ and $\lambda(q)$ given by
\beq
\l(q) = 1-q_M+\int_q^{q_M} \de p\, x(p)\:.
\label{def_lambda}
\eeq
Self-consistently, the functions $f(q, \underline h)$ and $P(q, \underline h)$ obey the $K$-dimensional partial differential equations: 
\begin{align}
\frac{\partial f(q,\underline h)}{\partial q} &= -\frac{1}{2}\left[\nabla_h^2 f(q,\underline h)+x(q) |\nabla_h f(q,\underline h)|^2 \right]
\label{full_correlated_f}
\end{align}
and
\begin{align}
\frac{\partial P(q,\underline h)}{\partial q}
&= \frac{1}{2}\left[\nabla_h^2 P(q,\underline h) -2x(q) \nabla_h \cdot \left(P(q,\underline h) \nabla_h f(q,\underline h) \right)\right]
\label{full_correlated_P} 
\end{align}
with the boundary conditions
\begin{align}
f(q_M,\underline h) &= 
\int d \underline{h} \frac{e^{-\frac{|\underline{h}|^2}{2(1-q_M)}}}{(2\pi(1-q_M))^{K/2}}
\theta(\Delta(\underline{h}))\\
P(q_m,\underline h) &= \frac{e^{-\frac{|\underline{h}|^2}{2 q_m}}}{(2\pi q_m)^{K/2}}.
\label{full_correlated_boundary_condition}
\end{align}


For later convenience, it is useful to present some additional identities. 
Within the interval of $x$ where $\dot q(x) \ne 0$, one may compute the first two derivatives of \cref{full_correlated} with respect to $q$, which yields
\beq
\begin{split}
	\frac{1}{\l^2(q)}&=\a \int_{-\infty}^\infty \de \underline h P(q,\underline h) \sum_{i,k=1}^K  \left(\partial_i\partial_k f(q,\underline h)\right)^2.\\
\end{split}
\label{replicone_correlated}
\eeq
and 
\begin{align}
x(q) = \frac{\lambda(q)}{2}  \frac{ \int_{-\infty}^\infty \de \underline h  P(q,\underline h) \sum_{i,j,k=1}^{K} (\partial_i \partial_j \partial_k f(q,\underline h)  )^2 }{\int_{-\infty}^\infty \de \underline h  P(q,\underline h)\SquareBracket{
		\sum_{i,j} (\partial_i \partial_j f(q,\underline h))^2  + \lambda(q) 	\sum_{i,j,k} (\partial_i \partial_j f(q,\underline h)) (\partial_j \partial_k f(q,\underline h)) (\partial_k \partial_i f(q,\underline h))
	}
}.
\label{breakingPoint}
\end{align}
These two relations, computed on the replica symmetric ansatz where $q(x)$ is a constant, give the replica symmetry breaking point and the value of the breaking point of $q(x)$ at the transition \cite{SD84, FPSUZ17}. 
Additionally, one may take another derivative w.r.t $q$ to compute the slope of $q(x)$ at the transition on the breaking point. 
The detailed steps to derive these equations can be found at the end of Sec.~\ref{sec:Derivation}, see also \cite{SD84, FPSUZ17}.

\begin{figure}[t]
	\centering
	\includegraphics[scale=0.7]{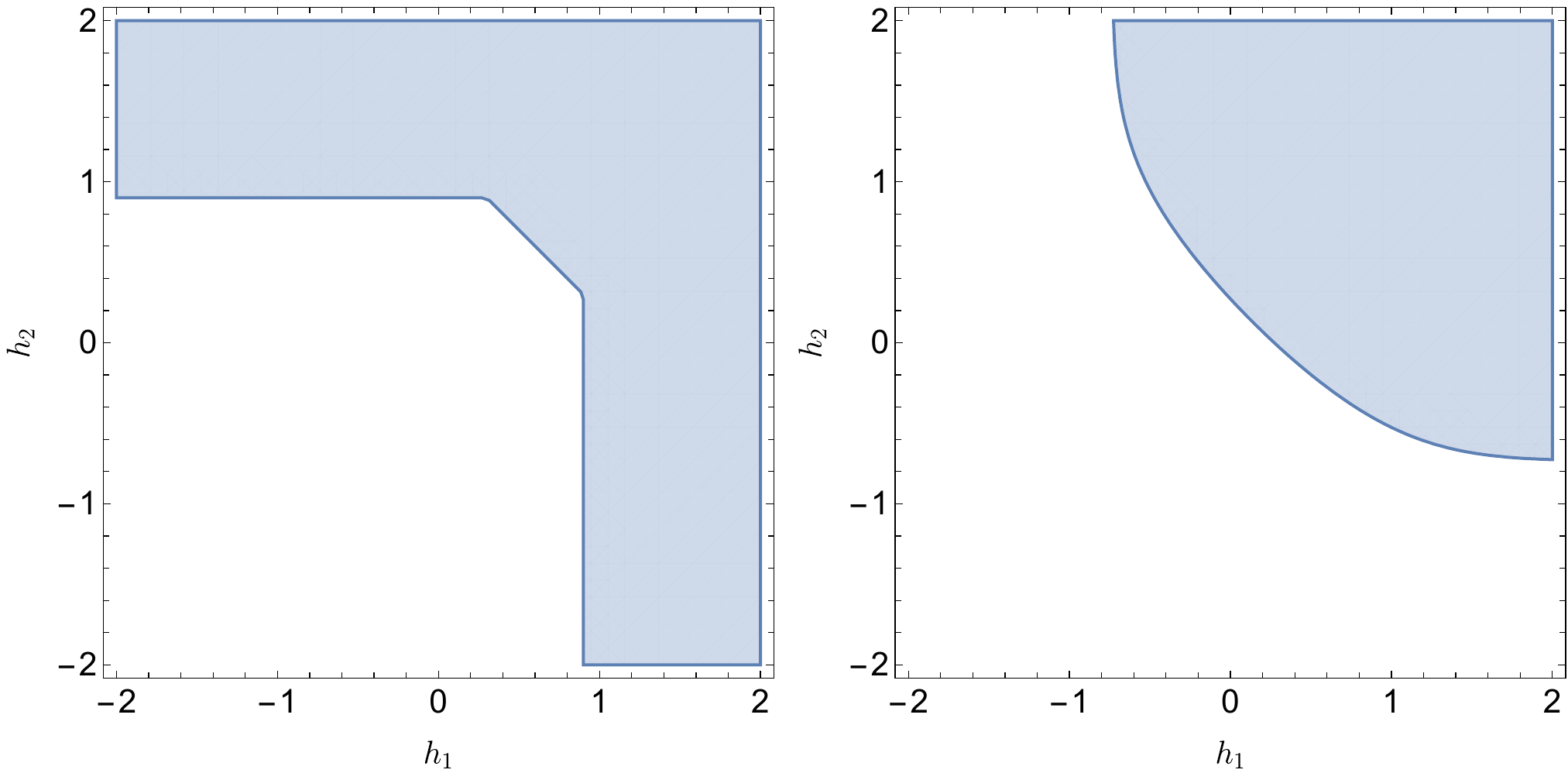}
	\caption{Plots of the allowed region $ \mathcal{R} = \{ \underline{h} \, | \, \D(\underline h) \geq 0\} $ for $K=2$  for (Left) the ReLU machine and (Right) the soft committee machine.  The parameter $\sigma$ is chosen to be $0.3$ in both cases.}
	\label{fig:allowedRegion}
\end{figure}

\section{Phase diagrams}
\label{sec:phase_diagrams}
Eqs.~(\ref{full_correlated_f}), (\ref{full_correlated_P}) and
(\ref{full_correlated}) contain all the information about our system. 
Here, we first study the phase diagrams to identify the fRSB phase.
In the simple perceptron \cite{FP15, FPUZ15, FPSUZ17}, it has been shown that this occurs in the non-convex regime.
Thus, it is important to make sure if this is also the case for the models we study in this paper.
In the simplest scenario, one can check whether the RS solution becomes unstable before it reaches the jamming transition, i.e., $q=1$. 
To this end, let us rewrite \cref{full_correlated,replicone_correlated} within the replica symmetric ansatz.
Setting $q(x) = q$  with $q_m = q_M =q$, these equations read
\begin{align}
\frac{q}{(1-q)^2} &=\a \int_{-\infty}^\infty\de \underline h \g^{(K)}_{q}(\underline h) |\nabla_h \ln  \Theta_{1-q}(\underline{h})|^2,
\label{full_correlated_RS}
\end{align}
and 
\begin{align}
\frac{1}{(1-q)^2} &= \alpha \int_{-\infty}^\infty d \underline h \gamma^{(K)}_{q}(\underline h) \sum_{i j}^K  \left(\partial_i \partial_j \ln \Theta_{1-q}(\underline{h})\right)^2,
\label{replicone_correlated_RS}
\end{align}
where 
\begin{align}
\Theta_{v}(\underline{h}) =  
\int d \underline{z} \frac{  e^{- \frac{|\underline{ h} - \underline{  z}|^2}{2v}} }{(2\pi v)^{K/2}} \theta(\underline{z} \in \mathcal{R}).
\label{ThetaFunction}
\end{align}
Note that the integral runs only over the allowed region $ \mathcal{R} = \{ \underline{h} \, | \, \D(\underline h) \geq 0\} $, specified by the model.
Specific examples of regions ${\cal R}$ for the ReLU and the soft-committee machines are shown in dark blue in \cref{fig:allowedRegion}.
Thus, for given $\alpha$ and the model $\mathcal{R}$, the order parameter $q$ satisfying either \cref{full_correlated_RS} or \cref{replicone_correlated_RS} can be computed numerically.
The so-called dAT line \cite{dAT78}, indicating the onset of instability of RS solution, is determined by finding $q$ that satisfies \cref{full_correlated_RS} and \cref{replicone_correlated_RS} simultaneously.
If there exists a solution for both equations at $q<1$, the RS solution becomes unstable before the jamming transition occurs.

The RS jamming line $ \curve{J} $ is determined by \cref{full_correlated_RS} at $q=1$.
By taking the limit $q\to1$, \cref{full_correlated_RS} is simplified to 
\begin{align}
1 &= \alpha \int_{-\infty}^\infty d \underline h \gamma^{(K)}_{1}(\underline h) 
\delta h_\perp^2 \theta(\underline{h} \notin \mathcal{R}),
\end{align}%
where $\delta h_{\perp}$ denotes the distance from $\underline h$ to the SAT-UNSAT boundary.

The point -if any- at which the RS jamming line and dAT line meet marks the border between 
RS and RSB jamming and can be determined by also 
considering the limiting behavior of \cref{replicone_correlated_RS} for $q\to 1$, assuming a finite limit we have:
\begin{align}
1 &=\alpha \int_{-\infty}^\infty d \underline h \gamma^{(K)}_{1}(\underline h) 
\theta(\underline{h} \notin \mathcal{R}).
\end{align}
Then this point is given by $\alpha$ for which both equations are satisfied.
We will show later with \cref{GapDistribution2}, the above condition implies the isostatic jamming at this point, i.e., $z =1$.

\begin{figure}[t]
	\centering
	\includegraphics[scale=0.7]{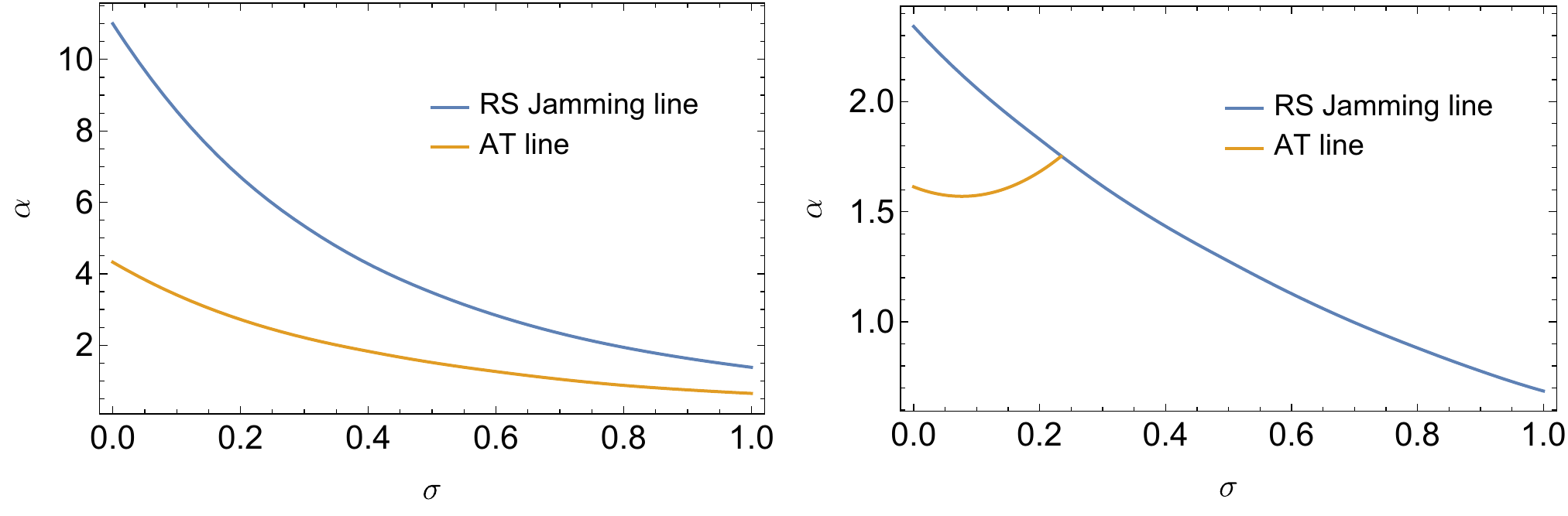}
	\caption{Sketch of the phase diagram of (Left) Relu $ (K=2) $ and (Right) soft-committee machine $ (K=3) $.
		This shows that there exists regions where the dAT line $\curve{dAT} $ is placed below the RS jamming line $\curve{J} $.
		In this case, the computation of breaking point \cref{breakingPoint} and its slope at $\curve{dAT} $ ensures that the jamming line lies in full-RSB phase.
		For ReLU, it can be shown that $\curve{J}$ is always above than $\curve{dAT}$ for all $\sigma >0$.	
	}
	\label{fig:phaseDiagram}
\end{figure}

This equation can only hold if for any point $\underline h$ the set of minimal distance on the RS SAT-UNSAT boundary is composed by a single point. 
In the parity and the ReLU machines this equation can never be satisfied, and the integral in \cref{replicone_correlated_RS} diverges as $ q \to 1 $, and the dAT line is always below the RS jamming line (See \cref{fig:phaseDiagram} (Left)). The jamming is isostatic and critical for all values of 
the model parameters. In the soft committee, instead, the equation is verified for ${\alpha_{rs}} \approx 1.5$ and  $\sigma_{rs} = 0.2$, in that case, above that values,  jamming is described by a RS solution hypostatic and non critical. 

In \cref{fig:phaseDiagram}, we have drawn these lines for (Left) ReLU $ (K=2) $ and (Right) soft-committee machine with $(K=3)$. 
From these two lines, we can identify the parameter range of $\sigma$ where the jamming transition occurs in the RSB phase. 
As a next step, we have to determine the type of RSB phase. We can proceed by determining the breaking point \cref{breakingPoint} and its slope at $ \curve{dAT} $ (See \cref{sec:Derivation}). 
In order to make a transition to full-RSB phase, the breaking point $m$ should be in the range $(0,1)$ and its slope be positive.
Within the range where the jamming occurs in the RSB phase, we confirmed for both models that  the breaking point and its slope satisfy these criteria, thus making a transition to fullRSB phase. 

However, this does not hold generally across all models. 
For example, it is known for Parity machine that the phase diagram undergoes first a RS-to-1RSB transition \cite{EVbook}. 
Even though we did not manage to verify the existence of the Gardner transition due to the increasing complexity of computation, we expect this happens given the clearly observed critical behaviors found in force and gap distributions from our simulation data.

\section{Scaling solutions near the SAT/UNSAT  transition}\label{sec:replicas}

The analysis of the
phase diagram in \cref{sec:phase_diagrams} shows that the assumption is justified
for all models at least in some ranges of the parameters. 
Here, we would like to study the jamming limit $q_M\to 1$ under the assumption of continuous RSB. 

\begin{figure}[t]
	\centering
	\includegraphics[scale=0.5]{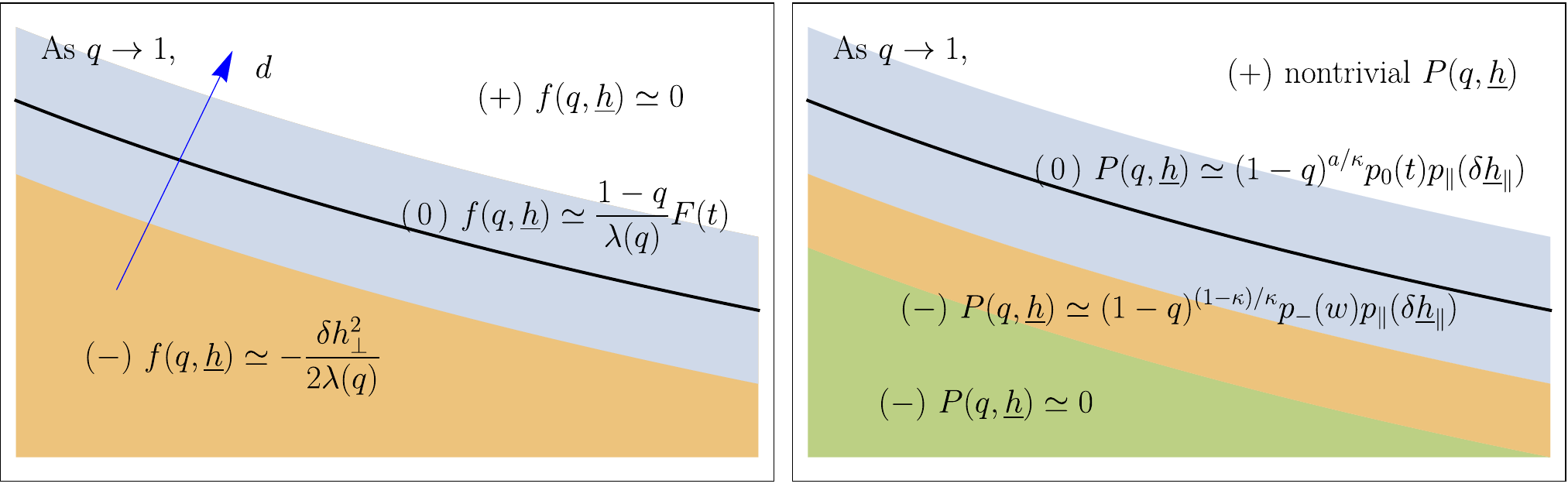}
	\caption{Schematic diagrams for different scaling regimes for $f(q, \underline{h})$ and $P(q, \underline{h})$. 
		The black thick line indicates the SAT-UNSAT boundary and the different shadings around this line indicate the corresponding scaling regimes.
		As a convention, the sign of $\delta h_{\perp}$ is chosen to be positive if $\underline{h} \in \mathcal{R}$. 
		The definition of the scaling functions $F(t)$, $p_0(t)$ are given in \cref{scalingF,scalingP}.
	}
	\label{fig:scaling_solution_schematic}\end{figure}

\subsubsection{Scaling solution of \cref{full_correlated_f}}
First of all, let us first examine the initial condition for $f(q,\underline{ h})$ given by \cref{full_correlated_boundary_condition}.
We will assume $\alpha < \alpha_J$ where the system approaches jamming from the SAT phase. 
As $q_M\to 1$, one can evaluate Eq.~(\ref{full_correlated_boundary_condition}) using the saddle point method.
This boundary condition depends on a single variable  $\delta h_{\perp}$, the distance to the SAT-UNSAT boundary delineated by the region $ \mathcal{R} = \{ \underline{h} \, | \, \D(\underline h) \geq 0\} $:
\begin{align}
f(q_M, \underline{h}) =  \left\{ \begin{array}{cc}
0 & \quad \text{i) $\underline{h} \in \mathcal R$}  \\ 
-\frac{\delta h_{\perp}^2}{2 (1-q_M)} & \quad \text{ii) $\underline{h} \notin \mathcal{R}$}  \\ 
\end{array}
\right..
\label{f_general_init}
\end{align}
For convenience, we additionally impose the sign of $\delta h_{\perp}$ such that  $ \delta h_{\perp} $ is positive (negative) if $ \underline{h} \in \mathcal{R} $ ($ \underline{h} \notin \mathcal{R} $).

Given the initial condition \cref{f_general_init}, the evolution of $f(q, \underline{h})$ with respect to $q$ can be computed via \cref{full_correlated_f}.
Since \cref{full_correlated_f} depends on a somewhat arbitrary function $x(q)$ at this point, it is crucial to make a reasonable ansatz for $x(q)$.
As we expect the singularity near $ q \sim 1 $ is developed at jamming, we employ the power-law ansatz, so that
\begin{align}
\frac{x(q)}{\lambda(q)} = - \frac{\lambda'(q)}{\lambda(q)} = \frac{\kappa-1}{\kappa} \frac{1}{1-q},
\label{ratioXtoLambda}
\end{align}
as $q\to 1$ with an unknown constant $\kappa$. The validity of this ansatz will be verified later.

Now, let us tackle \cref{full_correlated_f} within this ansatz. 
Close to jamming, the relevant configurations of fields ${\underline h}$ should lie
close to the border of the allowed region ${\cal R}=\{ {\underline h}|\Delta {(\underline h)}>0\}$.
For a given field $ \underline h $ close to the border, we denote the closest point on the border by $B$.
Setting the origin at $B$, we can parameterize
$ {\underline h}$ by the coordinate along the tangent plane on $B$,
$\delta {\underline h}_\parallel$ and the distance from it $\delta h_\perp$. 
Certainly, this parametrization is possible only when $\delta h_\perp \ll 1$, which is nevertheless the scaling regime of interest.
Depending on the scale of two length scales $\delta h_\perp$ and $\sqrt{1-q}$, we perform separate analysis for different scaling regimes.
\paragraph{Domain ($ + $) }
In this domain, we study the regime where $\delta h_{\perp} \gg \sqrt{1-q}$.
Since the initial condition $ f(q_M, \underline h) $ is zero everywhere, this holds also for any $q$:
\begin{align}
f(q, \underline h) =0.
\label{solutionFPositiveD}
\end{align}

\paragraph{Domain ($ - $)} 
Here, we study the regime where $\delta h_{\perp} \ll -\sqrt{1-q}$.
Since the transformation $\underline h \to (\delta h_{\perp}, \delta \underline {h}_\parallel )$ is locally a rotation: we have 
\begin{align}
|\nabla_h f|^2  = |\partial_{\delta h_{\perp}} f|^2 + |\nabla_{\delta \underline{h}_{\parallel}} f|^2. 
\end{align}
Since the boundary condition given by \cref{f_general_init} only
depends on $\delta h_{\perp}$, the leading singularity at $q=1$ only comes
from the contribution of $ |\nabla_{\delta h_{\perp}} f|^2 $. To the leading order in $(1-q)$ we have that:
\begin{align}
f(q, \underline h) = -\frac{\delta h_{\perp}^2}{2 \lambda(q)}(1 + O(\sqrt{1-q})).
\label{solutionFNegativeD}
\end{align}
By plugging this solution into \cref{full_correlated_f}, this solution can be easily verified.

\paragraph{ Domain (0) }
In this domain, we consider the matching regime $\delta h_{\perp} \sim  \sqrt{1-q} $ between the domain ($+$) and ($-$). 
In the derivation of \cref{solutionFNegativeD}, we showed that the Laplacian term in the (RHS) of \cref{full_correlated_f} is less singular compared to the non-linear term, 
which allows us to neglect this term.
However, when $\delta h_{\perp}$ becomes comparable to $ \sqrt{1-q} $, we are entering a new scaling regime in which a separate analysis is required.

To investigate this regime,  we introduce the scaling variable $t = \frac{\delta h_{\perp}}{\sqrt{1-q}}$. 
Then, \cref{full_correlated_f} is written in terms of new variables
$q, t$ and $ \delta \underline{h}_{\parallel}$, which we suppose not to scale with $1-q$. 
Additionally, we define $F(q,t, \delta \underline{h}_{\parallel}) =
\frac{\lambda(q)}{1-q} f(q, t, \delta \underline{h}_{\parallel}) $ that allows to
isolate the leading sinularity.  If we plug $F(q,t, \delta \underline{h}_{\parallel}) $
into \cref{full_correlated_f} the dependence
on $t$ provides singular terms, while the dependence on
$\delta \underline{h}_{\parallel}$ is regular and $F(q,t, \delta \underline{h}_{\parallel})$ is independent of $\delta \underline{h}_{\parallel}$ up to the leading order. 

After making the substitution, we find the  one-dimensional equation:
\begin{align}
\frac{t}{2} \frac{\partial}{\partial t} F(t)-\frac{F(t)}{\kappa } = - \frac{1}{2} \Paranthesis{
	\frac{\partial ^2}{\partial t^2 } F(t) + \frac{\kappa-1}{\kappa} \Paranthesis{
		\frac{\partial F(t)}{\partial t}
	}^2
} + O(\sqrt{1-q}).
\label{scalingF}
\end{align}
Imposing the matching condition on both sides with \cref{solutionFPositiveD,solutionFNegativeD}, the correct boundary conditions are determined to be
\begin{align}
F(t \to -\infty) = - \frac{t^2}{2},  \quad 
F(t \to 0) = 0.
\end{align} 
With these conditions, \cref{scalingF} is uniquely determined once $\kappa$ is given.
%

\subsubsection{Scaling solution of \cref{full_correlated_P}}
With the determination of $f(q, \underline{h})$, we are ready to evaluate \cref{full_correlated_P}.

\paragraph{ Domain ($+$)}
In this regime, it was derived in \cref{solutionFPositiveD} that $f(q, \underline h) = 0$. 
Thus, in the absence of $f(q, \underline{h})$ in \cref{full_correlated_P}, the equation is reduced to a $ K $-dimensional diffusion equation whose solution is readily obtained by a Gaussian profile:
\begin{align}
P(q, \underline{h}) = \frac{e^{-\frac{|\underline{h}|^2}{2 q}}}{(2\pi q)^{K/2}}.
\end{align}
for a given initial condition \cref{full_correlated_boundary_condition}.

\paragraph{Domain ($-$)}
In this regime, we can neglect the Laplacian term in \cref{full_correlated_P}. 
To see this, first notice that the initial condition for $ P(q,\underline h) $ in \cref{full_correlated_boundary_condition} is regular in $q$. Thus,
the singular behavior is developed at $q=1$ due to the singularities of $x(q)$ and $f(q, \underline h)$. 
With the Laplacian term dropped, \cref{full_correlated_P} is cast into a first-order partial differential equation:
\begin{align}
\frac{\partial P(q,\underline h)}{\partial q}
&= -x(q) P(q,\underline h)  \nabla_h^2 f(q,\underline h)
-x(q) \nabla_h f(q,\underline h) \cdot \nabla_h P(q,\underline h).
\end{align}
Plugging the solution \cref{solutionFNegativeD} of $f(q, \underline{h})$ and the power-law ansatz \cref{ratioXtoLambda} for $x(q)$ into the above equation, the leading behavior in $\delta h_\perp$ reads
\begin{align}
\frac{\partial P}{\partial q}
&= \frac{\kappa-1}{\kappa} \frac{1}{1-q} P 
+\frac{\kappa-1}{\kappa} \frac{\delta h_{\perp}}{1-q}   
\partial_{\delta h_{\perp}} P - \nabla_{ \delta \underline{h}_{\parallel}} f \cdot  \nabla_{\delta \underline{h}_{\parallel}} P + O(\delta h_{\perp}).
\end{align}
This equation can be understood using the method of characteristic. 
The first term in the (RHS) determines the overall growth of $P(q,\underline{h})$ given by the growth equation:
\begin{align}
\frac{dP}{P} = \frac{\kappa-1}{\kappa} \frac{dq}{1-q},
\end{align}
which gives $ P \propto (1-q)^{\frac{1-\kappa}{\kappa}}  $.
On the other hand, the other terms dictate the $K$ conserved quantities along the characteristic manifold.
Let us compute the conserved quantity $w$ associated to $\delta h_\perp$, which describes the change of $ \delta h_\perp$ due to the change of $q$:
\begin{align}
\frac{d \delta h_\perp}{d q} = \frac{\kappa-1}{\kappa} \frac{\delta h_{\perp}}{1-q}.
\end{align}
The solution of this one-dimensional ODE readily follows $ w = \delta h_{\perp} (1-q)^{\frac{1-\kappa}{\kappa}}  $.

Finally, one can in principal determine the remaining $K-1$ conserved quantities following the same procedure.  
However, this would be impossible since it requires the full knowledge of $ \nabla_{ \delta \underline{h}_{\parallel}} f $.
Nevertheless, we can claim that these conserved quantities are regular (or at least less singular than $w$) from the solution \cref{solutionFNegativeD}. 

Combining all these findings together, $ P(q, \underline{h}) $ is of the form:
\begin{align}
P(q, \underline{h}) \sim (1-q)^{\frac{1-\kappa}{\kappa}} p_-(w)p_{\parallel}(\delta \underline{h}_{\parallel}, q) 
\label{SolutionP}
\end{align}
where $p_-$ and $p_\parallel$ are determined from the profiles at a fixed $q$.

There are two important consequences from this solution.
First of all, we have just shown that the probability measure should be concentrated within the scale $ w\sim O(1)$. 
Second, the factorization property of this solution implies that along the perpendicular direction to border, our equation is completely reduced to the one-dimensional Parisi equation.

\paragraph{Domain (0)} 
In this scaling regime where $ \sqrt{1-q} \sim \delta h_\perp$, two terms of (RHS) in \cref{full_correlated_P} are properly balanced. 
Following the same approach for $f$ as in \cref{scalingF}, we want to rewrite $P(q, \underline{h})$ in terms of the new scaling variable: $ (q, \underline h) \to (q,t, \delta \underline{h}_{\parallel}) $. 
Introducing a function $p_0(t)$ such that $ P = (1-q)^{-a/\kappa} p_0(t) p_{\perp}(\delta \underline{h}_{\parallel}) $, we obtain
\begin{align}
\frac{a}{\kappa} p_0(t) + \frac{1}{2} t p_0'(t)  =  \frac{1}{2} p_0''(t)  - \frac{\kappa-1}{\kappa} (p_0(t) \mathcal{F}'(t))' + O(\sqrt{1-q}),
\label{scalingP}
\end{align}
where the prime denotes the derivative with respect to $t$. 
As previously shown for $F(t)$, it is again reduced to a one-dimensional equation.
Finally, imposing the smooth matching conditions on both sides, one arrives at the boundary conditions
\begin{align}
p_0(t \to \infty ) = t^{-\gamma}
\end{align}
where $ \theta = \frac{2a}{\kappa} $
and 
\begin{align}
p_0( t \to -\infty ) = |t|^{\theta},
\end{align}
where $ \theta = \frac{1- \kappa +a}{\kappa/2 - 1} $.
Given $\kappa$, there exists a unique $a$ that satisfies both boundary conditions by which $a$ is determined.

Finally, it is worth to remark that, via the relation $ \mathcal{M}(t) = - F'(t)$, \cref{scalingF,scalingP} are equivalent to the scaling functions found in \cite{FPSUZ17}.
This implies that regardless of the choice of the models, the solutions $F(t)$ and $p_0(t)$ are determined independently by the same equation for the standard perceptron model.

\subsubsection{Determination of $\kappa$}
We have shown that for a given $\kappa$, the scaling functions $F(t)$ and $p_0(t)$ are uniquely determined. 
Moreover, they follow the same equations as the ones for the standard perceptron.
The only task still needs to be verified is to check if $\kappa$ is the same as for the case of standard perceptron.
It can be achieved by evaluating \cref{breakingPoint}.
By dividing both sides by $\lambda(q)$ in \cref{ratioXtoLambda},
$\kappa$ should verify the following equation:
\begin{align}
\frac{\kappa - 1}{\kappa} = \frac{1}{2} \frac{ \int_{-\infty}^\infty \de \underline h  P(q,\underline h) \sum_{i,j,k} (\partial_i \partial_j \partial_k f(q,\underline h)  )^2 }{\int_{-\infty}^\infty \de \underline h  P(q,\underline h)\SquareBracket{
		\sum_{i,j} (\partial_i \partial_j f(q,\underline h))^2  + \lambda(q) 	\sum_{i,j,k} (\partial_i \partial_j f(q,\underline h)) (\partial_j \partial_k f(q,\underline h)) (\partial_k \partial_i f(q,\underline h))
	}
}
\label{ratioXtoLambda2}.
\end{align}
This integral can be evaluated by splitting the integral domain into different scaling regimes. 
By examining the behaviors of $f(q,\underline{h})$ and $P(q, \underline{h})$ sketched in \cref{fig:scaling_solution_schematic}, one can check that the only non-zero contribution comes from the scaling regime (II) in \cref{fig:scaling_solution_schematic} near the boundary, i.e., 
\begin{align}
\frac{\kappa - 1}{\kappa} = \frac{1}{2} \frac{ \int \de t  p_0(t)  F'''(t)^2 }{\int \de t  p_0(t) F''(t)^2  \SquareBracket{ 1 -F''(t)  
	}
}.
\end{align}
After the substitution $F'(t) = -\mathcal{M}(t)$, we recover the same equation for the standard perceptron.

\subsubsection{Wedge domain}
So far, we have analyzed \cref{full_correlated_f,full_correlated_P} near the SAT-UNSAT boundary locally allowing a transformation along the SAT-UNSAT boundary. 
Although this condition holds for most points in various models, there are examples that it is not always the case.
For example, the allowed region in \cref{fig:allowedRegion} (Left) contains two non-differentiable vertices at the boundary. 
Near one of them $(h_{1v}, h_{2v})$, we need to introduce additional scaling regime where two coordinates from the boundary are both small. 
Specifically, by setting $(t_1, t_2) = \Paranthesis{\frac{h_1 - h_{1v}}{\sqrt{1-q}}, \frac{h_2 - h_{2v}}{\sqrt{1-q}} }$, one can derive from \cref{full_correlated_f} a non-trivial two-dimensional equation 
\begin{align}
-\frac{F(t_1,t_2)}{\kappa } + \sum_{i=1}^2 \frac{t_i}{2} \frac{\partial}{\partial t_i} F(t_1,t_2 ) = - \frac{1}{2}  \sum_{i=1}^2\Paranthesis{
	\frac{\partial ^2}{\partial t_i^2 } F(t_1,t_2) + \frac{\kappa-1}{\kappa} \Paranthesis{
		\frac{\partial F(t_1,t_2)}{\partial t_i}
	}^2
} + O(\sqrt{1-q}),
\label{scalingF2Dim}
\end{align}
where the boundary condition is given in terms of the opening angle $\theta$ of the vertex.
As a result, the solution in general depends on $\theta$, and thus be non-universal.
This implies that if we collect only the gap variables $\underline{h}$ sufficiently near the wedge domain, the statistical properties is no longer universal.

With this non-universal behavior, one may further ask if the exponent $\kappa$ is affected by these points.
This question can be answered by reexamining \cref{breakingPoint} in this wedge domain. 
However, one can immediately see that with one additional factor $\sqrt{1-q}$ due to a double scaling $(t_1, t_2) = \Paranthesis{\frac{h_1 - h_{1v}}{\sqrt{1-q}}, \frac{h_2 - h_{2v}}{\sqrt{1-q}} }$, the contribution of this scaling regime vanishes quickly as $\sqrt{1-q}$. 
Consequently, even in the presence of non-differentiable points at the SAT-UNSAT boundary, the universal power-law exponent $\kappa$ is unchanged.

\section{Force and gap distributions at jamming transition}
So far, we have focused on constructing the scaling behavior upon approaching jamming from SAT phase. 
However, to compute the statistics of forces, and contacts near jamming, it is important also to consider the scaling theory in the UNSAT phase. 
Along the same lines used in \cite{FPSUZ17}, we can introduce a joint probability distribution of the variables $\underline h$. Its expression $\rho(\underline h)$ can be computed from replicas as in \cite{FPSUZ17} for a finite inverse temperature $\beta$:
\begin{align}
\rho(\underline h) = e^{-\beta v(\underline h)} \int d\underline z \, P(q_M, \underline z) e^{-f(q_M, \underline z)} \frac{e^{-\frac{|\underline z - \underline h|^2}{2 (1-q_M)}}}{(2\pi (1-q_M)^{K/2}},
\label{GapDistribution}
\end{align}
where the potential is given by $ v(\underline{h}) = \frac{1}{2} \delta h_{\perp}^2 \theta( -\delta h_{\perp})$.
In the zero temperature limit, the product $\beta(1-q_M)$ converges to a fixed value $ \chi$.  
Then, the matching condition with SAT solution enforces $\chi$ to diverge upon approaching the jamming transition.

First, let us compute $ f(q_M, \underline h) $ in the UNSAT phase, which now gains additional contributions from the potential energy. Following the same procedures as in \cref{f_general_init}, the solution is recast into a minimization problem:
\begin{align}
f(\chi, \underline{h}) = \beta \min_{\underline z \notin \mathcal{R}} \SquareBracket{
	-\frac{|\underline h - \underline z|^2}{2 \chi} - \frac{\delta h_{\perp}^2}{2}
} \theta(\underline h \notin \mathcal{R}).
\label{f_general_init_UNSAT}
\end{align}

In the limit $\chi \to \infty$, the first term becomes  sub-dominant and the minimization for the second term implies that $\delta h_{\perp}$ should be of the order $O(\chi^{-1})$. 
Additionally, we establish that the nearest point at the boundary for $\underline h$ and $\underline{z}$ differ only by the amount of the order $O(\chi^{-1})$.
These observations lead us to find the solution:
\begin{align}
f(q_M, \underline{h}) = - \beta \frac{\delta h_{\perp}^2}{2\chi(1+ O(\chi^{-1})) }.
\label{f_general_init_UNSAT2}
\end{align}
This equation has two important implications. 
First of all, Eq.~(\ref{GapDistribution}) becomes
\begin{align}
\rho(\underline h) = P(1, \underline h (\chi + o(1)  )) (\chi + o(1) \theta(\underline h \notin \mathcal R) + P(1 , \underline h) \theta(\underline h \in \mathcal R).
\label{GapDistribution2}
\end{align}
From this, we can compute both the force and gap distributions. 
Indeed it follows that up to rescaling by a constant factor, $\rho(\underline h)$ has a trivial relation with $P(1, \underline h) $.
Using this we can show that the force and gap distributions have the power law behaviors described in the main text. 

Second, \cref{replicone_correlated} becomes
\begin{align}
1 = \alpha \int d \underline{h} \, P(1, \underline h) \theta(\underline h \notin \mathcal{R}).
\end{align}
Thus \cref{GapDistribution2} implies that the number of contacts is $ N $ at jamming which means that the system is \emph{isostatic}.

Having determined the distribution $\rho(\underline h)$, we may now compute the distribution of gaps. 
By definition, it is given by
\begin{align}
\rho(\Delta) = \int d\underline h \, \rho(\underline h)  \delta( \Delta - \Delta(\underline h) ).
\end{align}
Since the contours given by the condition $\delta h_{\perp} =0$ and $\Delta(\underline h) =0$ coincide, 
they have locally a linear relationship, i.e., 
$ \delta h_{\perp} \simeq \Delta(\underline h) / | \nabla_h \Delta(\underline{h}) | + o(\Delta) $.
Also, using our scaling solution \cref{SolutionP}, we have the following
\begin{align}
\rho(\Delta) &\stackrel{\Delta \to 0^+}{=} \Delta^{-\gamma}  \int d\underline h \,  |\nabla_h \Delta(\underline{h}) |^{\gamma} p_{\perp}(\delta \underline{h}_{\parallel}(\underline h))   \delta( \Delta - \Delta(\underline h) ) \nonumber\\
&= \Delta^{-\gamma} (\mathrm {const}  + O(\Delta)  ).
\end{align}
This implies that the gap distribution also follows the power-law with the same exponent.

\section{Simulation details and results}
\begin{figure}[t]
	\centering
	\includegraphics[scale=0.7]{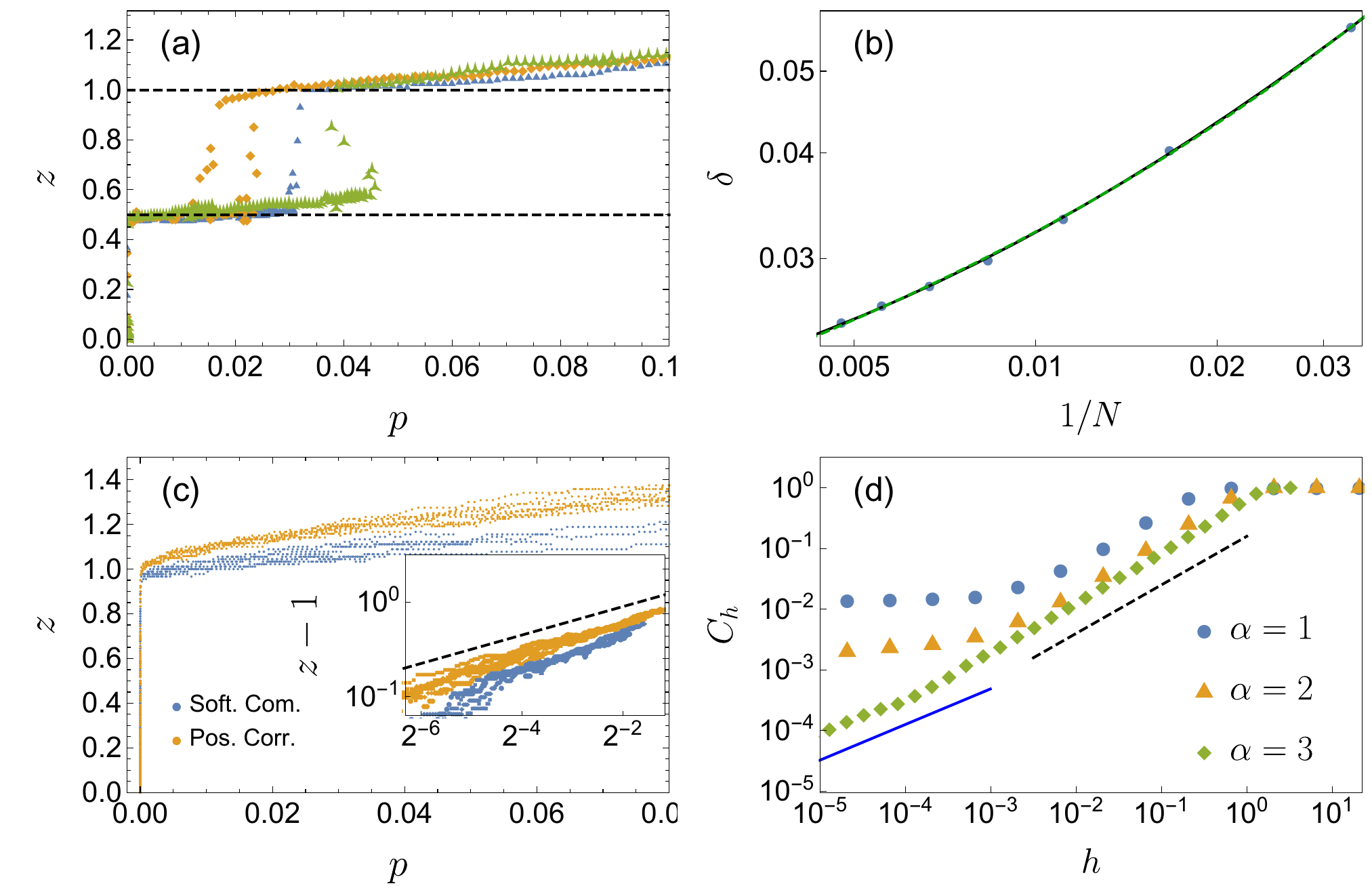}
	\caption{(a) Contact number as a function of pressure for three typical realizations for ReLU with $K=2$ and $N=240$. The curve can be roughly characterized by two jumps each with height roughly $1/2$. Due to the finite size effect, the position of the second jump fluctuates strongly in the range of system sizes we checked, e.g., $N < 500$.
		(b) Double-logarithmic plot of the average interval $\delta$ between two jumps as a function of $N^{-1}$. As $N$ increases, $\delta$ is expected to decrease to zero. 
		The green solid line and the black dashed line correspond to the best fits for two three parameter models: i) $ A x^\beta + B $ and ii) $A x^\beta + B x$, respectively. 
		The fitted parameters are i) $ A = 0.499, B = 0.015 $ and $\beta = 0.733$ and ii) $ A = 0.06, B=0.758 $ and $\beta =0.19$. 
		Given the nearly perfect fits for both models, it is not conclusive solely from the fits whether the curve converges to zero.
		(c) Contact number as a function of pressure for three typical realizations for soft-committee machine with $K=2$ and $N=240$ and the correlated perceptron with $\rho = -0.5$ and $N=240$. 
		Up to statistical noise, the two main consequences of our scaling theory are corroborated, i) the contact reach a isostatic point at jamming and ii) $ z-1 \sim p^{1/2}$.
		(d) Positive gap distribution of soft-committee machine for various $\alpha$ with $N=240$ and $K=3$. 
		As decreasing $\alpha$, the system moves from the full-RSB phase to RS phase.
		Accordingly, the critical range of gap variables seems to shrink as $\alpha$ passes through the RS-RSB boundary.
	}
	\label{fig:FiniteSizeEffect}
\end{figure}

As reported in the main text, the jamming point for this class of non-overlapping multilayer machines may not be well-defined due to a finite-size effect. 
Especially, we have noticed that for the case of the ReLU machine the $K$-units behave quasi-independently, thus the number of contacts exhibits a multiple jumps rather than one jump. 
This is due the finite $N$ sample-to-sample fluctuations of the jamming point for each effective field.
Since these fields are statistically equivalent, we expect these jumps will be concentrated into a single point in the infinite size limit. 
In \cref{fig:FiniteSizeEffect} (a) and (b), we present the typical fluctuation observed in the simulation. In \cref{fig:FiniteSizeEffect} (a), we show three typical behaviors of contact number $z$ as a function of pressure $p$. This illustrates that it is roughly characterized by two jumps and the first one appears at the isostastic point. 
However, the position of second jump fluctuates strongly for the system sizes available in the simulation $N < 500$. 
To show numerically that this fluctuation decreases in the infinite size limit, the average interval $\delta$ between two jumps as a function of inverse system size $N^{-1}$ is shown in a double logarithmic scale (\cref{fig:FiniteSizeEffect} (b)). 
The curve exhibits a clear decreasing behavior (to the left in the figure) which might be characterized by a power-law decay. 
However, within the parameter range we can control in our simulation, we cannot obtain a conclusive result regarding whether it definitely converges to zero or to a non-zero constant. 
Our fitting results based on the two three-parameter models, i) $ A x^\beta + B $ and ii) $A x^\beta + B x$, imply that the intercept, if it is not non-zero, should be small.

For completeness, we have drawn the similar $ z $ vs $ p $ curves for the soft committee machine and the correlated perceptron in \cref{fig:FiniteSizeEffect} (c).
We find that: i) $z$ reaches an isostastic point at jamming and ii) $z-1 \sim p^{1/2}$.

Finally in \cref{fig:FiniteSizeEffect} (d), we present the positive gap distribution upon approaching the SAT phase by varying $\alpha$ for the soft-committee machine. 
For large $\alpha$, we have noticed a cross-over behavior characterized by two power-law exponents from $1-\gamma$ (solid line) to another exponent approximately $1 - 0.2$ (dashed line). 
A similar crossover behavior is observed also in the ReLU machine. 
On the other hand, if $\alpha = 1.0$, the system is in the RS phase, thus we do not expect the gap distribution to follow the same power-law exponent predicted by our scaling theory.
Also, the jamming is observed to be hypostatic characterized by the condition $ z<1 $.

Now, let us turn our attention to the intermediate case $ \alpha =2 $.
According to \cref{fig:phaseDiagram}, the system is located near the RS-RSB boundary.
Even though the system is technically in the full-RSB phase, our simulation results suggest that the gap distribution does not quite achieve the expected power-law distribution. 
However, we attribute this to the finite size effects; near the boundary, some samples behave as if they are in RS phase, yielding different statistics in the gap distribution.

\section{The Teacher-Student setting}
So far we have considered the problem of storing random patterns in simple 2-Layers machines.
However, networks trained in this way cannot generalize, just because the training set is pure noise. 
In order to consider the generalization properties of these networks we need to consider the situation in which the patterns have a structure.
A simple setting in which this can be studied is the Teacher-Student one. 
In this case a machine, the Teacher, built from random weights $\doubleunderline w^*$ generates a set of input-output associations by computing
the output of random inputs. This training set is then learned by the Student, a network whose architecture is the same as the Teacher's. 
Therefore the learning problem becomes an inference problem for the weights matrix $\doubleunderline w$ \cite{KZ16, AMBKMZ18}. 
In order to fix the ideas we consider a parity machine with $K\geq 3$. In this case, one can show that for small $\a$, again the size of the training set, generalization is impossible,
since the typical overlap $R=\underline w_i\cdot \underline w_i^*$ between the weights of the Teacher and the Student is zero \footnote{Note that in principle $R$ depends on the unit $i$ but thanks to the statistical equivalence between hidden units such dependence disappears.}. 
Increasing $\a$, a first order phase transition happens at $\a_{\textrm K}$ and $R$ jumps to a finite value:  here the Student starts to generalize.
This first order transition is only theoretically possible. 
The solution $R=0$ continues to exist for $\a>\a_{\rm K}$ but here it is metastable.
One can show that the landscape of the free energy restricted to this metastable branch is glassy \cite{SST92, AFUZ18}
and one can follow the evolution of glassy states increasing $\a$. 
If $\s=0$ all glassy states are stable at arbitrarily large values of $\a$. 
Therefore one can expect that trying to learn $\doubleunderline w$ with local algorithms leads to glassy configurations in the metastable branch \cite{AFUZ18}.
These glassy states may jam eventually at $\a>\a_K$ and there one expects the same critical exponents of the purely random model. 
On the contrary, if $\s>0$ glassy states may undergo a spinodal transition towards the stable branch at $R>0$ (a kind of Kirkwood instability \cite{FP87, FP99, AFUZ18}) 
and one can have that either the instability arrives before jamming
or vice versa. Again, in the latter case one expects jamming to be in the same universality class as hard spheres.
Therefore under Bayes-optimal conditions in which the Student knows perfectly the probabilistic nature of the Teacher, jamming appears as a point in 
the metastable branch where the Student is learning by heart.
In the phase where perfect generalization is achieved instead, jamming is never found for finite $\a$ and continuous weights.
This scenario changes as soon as the Student does not have the complete probabilistic information on the Teacher.
However one can show in simple models that out-of-Bayes-optimality learning curves typically display over-fitting near jamming \cite{EVbook}.
This is reasonable since there one is supposed to learn perfectly the training set in the most compressed way.
Therefore good learning machines need to be far away from jamming.

\section{Determining the breaking point and its slope}
\label{sec:Derivation}
In this section, we derive \cref{replicone_correlated,breakingPoint} and also its slope $x'(q)$ from \cref{full_correlated} by taking a successive derivatives with respect to $q$.
For the sake of simplicity, we omit the arguments of $P(q, \underline h)$ and $f(q, \underline h)$ in the following.

First, let us define the following quantities:
\begin{align}
&A_m = \sum_{i_1, \cdots, i_m} \frac{d}{dq} \int_{-\infty}^\infty d \underline h P \left( f_{i_1, \cdots, i_m}\right)^2 \nonumber \\
&= \sum_{i_1, \cdots, i_m} \int_{-\infty}^\infty d \underline h \dot{P} \left(   f_{i_1, \cdots, i_m}\right)^2
+ \sum_{i_1, \cdots, i_m} 2 \int_{-\infty}^\infty d \underline h P f_{i_1, \cdots, i_m} \dot{f}_{i_1, \cdots, i_m},
\end{align}
where 
$f_{i_1, \cdots, i_m} =  \Paranthesis{
	(\partial_{i_1} \cdots \partial_{i_m} )} f $.

Using \cref{full_correlated_P} and performing integration by parts repeatedly for $P(q, \underline h)$, we obtain
\begin{align}
A_m =& \sum_{i_1, \cdots, i_m,k } \frac{1}{2} \int_{-\infty}^\infty d \underline h P  \Paranthesis{
	\partial_k^2 + 2 x (\partial_k f) \partial_k 
} f_{i_1, \cdots, i_m}^2 \nonumber\\
&- \sum_{i_1, \cdots, i_m,k}  \int_{-\infty}^\infty d \underline h P f_{i_1, \cdots, i_m} \Paranthesis{
	(\partial_{i_1} \cdots \partial_{i_m} ) \partial_k^2 f + x 	(\partial_{i_1} \cdots \partial_{i_m} ) (\partial_k f)^2
}
\end{align}
Then, $A_m$ can be further simplified with the identity
\begin{align}
\frac{1}{2} \partial_k^2 f_{i_1, \cdots, i_m}^2 - f_{i_1, \cdots, i_m} (\partial_{i_1} \cdots \partial_{i_m} ) \partial_k^2 f = f_{i_1, \cdots, i_m, k }^2,
\end{align}
leading us to find
\begin{align}
A_m =& \sum_{i_1, \cdots, i_m,k} \int_{-\infty}^\infty d \underline h P \Paranthesis{
	f_{i_1, \cdots, i_m, k }^2 + x (\partial_k f) \partial_k 
	f_{i_1, \cdots, i_m}^2 
	-   x f_{i_1, \cdots, i_m}	(\partial_{i_1} \cdots \partial_{i_m} ) (\partial_k f)^2
}.
\end{align}
We need this expression for $m=1,2,3$, which reads
\begin{align}
A_1 &= \sum_{i,j} \int_{-\infty}^\infty d \underline h P 
f_{i, j }^2  = \Avr{f_{i, j }^2} \nonumber \\
A_2 &= 	\Avr{
	f_{i, j, k }^2	- 2 x f_{ij} f_{jk}f_{ki}
} \nonumber \\
A_3 &= \Avr{
	f_{i, j, k, l}^2	- 2 x f_{ijk} \Paranthesis{
		f_{il} f_{jkl} 
		+ f_{jl} f_{kil} 
		+ f_{kl} f_{ijl} 
	} 
} \nonumber\\
&= \Avr{
	f_{i, j, k, l}^2	- 6 x f_{ijk}
	f_{il} f_{jkl} 
}.
\label{EqA}
\end{align}
Here, we use the bracket notation to denote the combined action of the integral over $\underline h$ with density $P(q,\underline h)$ and the summation over all indices.

Similarly, let us evaluate the following quantity
\begin{align}
B =& \frac{d}{dq} \Avr{ f_{ij} f_{jk} f_{ki} }
= \sum_{i,j,k} \int_{-\infty}^\infty d \underline h \dot{P} f_{ij} f_{jk} f_{ki}
+\sum_{i,j,k}3  \int_{-\infty}^\infty d \underline h P \Paranthesis{
	\dot{f}_{ij} f_{jk} f_{ki}
}\nonumber \\
=& 3 \Avr{
	f_{ijl} f_{jkl} f_{ki} 
	- x f_{jk}f_{ki}f_{il}f_{jl}
}
\end{align}
where we simplify the terms by rearranging indices:
\begin{align}
\partial_l f_{ij} f_{jk} f_{ki} &= 3 f_{ijl} f_{jk} f_{ki} \nonumber \\
\partial_l^2 f_{ij} f_{jk} f_{ki} &= 3 f_{ijll} f_{jk} f_{ki}
+ 6 f_{ijl} f_{jkl} f_{ki}.
\end{align}
To derive \cref{replicone_correlated}, we differentiate the both sides of \cref{full_correlated} with respect to $q$. 
Using \cref{EqA}, this reads
\begin{align}
\frac{1}{\lambda^2} = \alpha 		A_1,
\end{align}
which is just a short-form of \cref{replicone_correlated}.

Next, the equation for breaking point \cref{breakingPoint} may be derived similarly. 
After multiplying the both sides of \cref{replicone_correlated} by $\lambda^2 /\alpha $ and further taking the derivative with respect to $q$, we find
\begin{align}
0 &= - 2 x A_1  + \lambda A_2 \nonumber\\
&= - 2 x A_1  + \lambda \Avr{f_{ijk}^2} - 2 \lambda x \Avr{f_{ij} f_{jk} f_{ki} }.
\label{breakingPointDetail}
\end{align}

By rearranging the terms for $x$, we find 
\begin{align}
x = \frac{\lambda \Avr{f_{ijk}^2}}{2 (A_1 + \lambda  \Avr{f_{ij} f_{jk} f_{ki} })},
\end{align}
which is again the shorthand form of \cref{breakingPoint}.

Finally, let us take another derivative of \cref{breakingPointDetail} with respect to $q$. 
Then, the terms are expanded in the following way:
\begin{align}
0 &= - 2 x' A_1 - 2 x A_2  - x A_2 + \lambda \partial_q A_2  \nonumber\\
&= -2 x' A_1 - 3 x A_2 + \lambda ( A_3 - 2 x' \Avr{f_{ij} f_{jk} f_{ki}} - 2x B )  \nonumber \\
2x'(A_1 + \lambda \Avr{f_{ij} f_{jk} f_{ki}})  &= -3x A_2 + \lambda A,
\end{align}
where 
\begin{align}
A &= A_3 - 2x B \nonumber\\
&=\Avr{f_{i, j, k, l}^2} - 6 x \Avr{f_{ijk} f_{il} f_{jkl} } - 6 x \Paranthesis{
	\Avr{ f_{ijl} f_{jkl} f_{ki} }
	- x \Avr{f_{jk}f_{ki}f_{il}f_{jl}}
} \nonumber \\
&= \Avr{f_{i, j, k, l}^2} - 12 x \Avr{f_{ijk} f_{il} f_{jkl} } + 6 x^2 \Avr{f_{jk}f_{ki}f_{il}f_{jl}}.
\end{align}
This can be further simplified using $A_1 + \lambda \Avr{f_{ij} f_{jk} f_{ki}} = \lambda \Avr{f_{ijk}^2}/(2 x)  $ and $A_2 = 2x A_1 /\lambda$:
\begin{align}
2 x' \frac{\lambda \Avr{f_{ijk}^2}}{2x} = - 3 x \frac{2x A_1}{\lambda} + \lambda A \end{align}
which then gives
\begin{align}
x' = \frac{2x}{\alpha \Avr{f_{ijk}^2}} \Paranthesis{
	\frac{A \alpha}{2} - \frac{3x^2}{\lambda^4}
},
\end{align}
where we used $A_1 = (\alpha \lambda^2)^{-1}$.

\bibliography{HS}